# Giant magnetothermopower in charge ordered $Nd_{0.75}Na_{0.25}MnO_3$

D. V. Maheswar Repaka and R. Mahendiran[1]

Department of Physics, 2 Science Drive 3, Faculty of Science,

National University of Singapore, Singapore 117551, Singapore

We report magnetization, resistivity and thermopower in the charge-orbital ordered antiferromagnet $Nd_{0.75}Na_{0.25}MnO_3$. Magnetic-field induced collapse of antiferromagnetism is found to be accompanied by a giant negative magnetothermopower (= 80-100% for a field change of 5T) over a wide temperature ($T$ = 60-225K) and giant magnetoresistance. While the field-induced metamagnetic transition in magnetization is reversible upon field-cycling at $T > 40$ K, it is irreversible at lower temperatures and this has impact on magnetoresistance, magnetothermopower as well as change in the temperature of the sample. Our results indicate high sensitivity of thermopower to changes in the magnetic state of the sample.

---

[1] Corresponding author (phyrm@nus.edu.sg)

Thermoelectric materials exploiting the Seebeck effect can convert waste heat into electrical voltage. The Seebeck coefficient (thermopower) is defined as $\alpha = \Delta V/\Delta T$ where $\Delta T$ is the temperature difference between the hot and cold ends of a sample and $\Delta V$ is the electrical potential difference created. In recent years, urge to understand and exploit interplay between spin and thermal transport is growing following the discovery of spin-dependent Seebeck effect in giant magnetoresistive granular alloys[1,2] and magnetic tunnel junctions.[3] The relative change in the magnetization directions of two ferromagnetic electrodes in these nanostructures not only affects electrical resistance but also influences thermopower.[4] Magnetothermopower (MTEP) is spin-dependent in these magnetic nanostructures and it arises from differences in the Seebeck coefficients for spin up and spin down electrons in the ferromagnetic electrode. Spin-dependent Seebeck effect can occur also in bulk materials such as perovskite manganites in which electrical transport is strongly spin-dependent since hopping of $e_g$ hole between $Mn^{4+}:t_{2g}^3 e_g^0$ and $Mn^{3+}:t_{2g}^3 e_g^1$ is enhanced if $t_{2g}^3$ core spins on both the sites are parallel and prohibited if they are antiparallel. Although magnetoresistance(MR) has been extensively investigated in manganites over the past two decades, interplay between magnetism and thermopower has been scarcely investigated so far. Until now, there are only a very few studies of MTEP in manganites and they are confined to compositions exhibiting paramagnetic- ferromagnetic transition.[5,6,7,8,9]

In this report, we investigate correlations between electrical resistivity, magnetization and magnetothermopower in the charge-ordered antiferromagnetic insulator $Nd_{0.75}Na_{0.25}MnO_3$. The substitution of $x$ number of monovalent Na in $NdMnO_3$ introduces $2x$ numbers of holes ($Mn^{4+}$ ion) in the $Mn^{3+}$ matrix. The A-type antiferromagnetic $NdMnO_3$ turns into an inhomogeneous ferromagnet for $x$ = 0.1-0.15.[10] For the nominal composition x = 0.25, numbers of electrons and holes are equal ($Mn^{3+}:Mn^{4+}$ = 1:1) and they order in a checker board configuration below the charge-orbital ordering temperature $T_{CO}$ = 180K followed by antiferromagnetic ordering at a lower temperature $T_N$ = 120K. Magnetothermoelectric response of this composition is not reported so far.

Polycrystalline $Nd_{0.75}Na_{0.25}MnO_3$ sample was prepared using polymer assisted sol-gel method using metal nitrates as starting precursor materials. Details of the sample preparation is beyond the scope of this paper. Magnetization, MR and MTEP were measured using a commercial cryostat and a vibrating sample magnetometer. In the zero-field cooling (ZFC) mode, $M(T)$ was recorded while warming in a specific field after cooling the sample in zero field to 10K. In the field-cooled cooling (FCC) and warming (FCW) modes, $M(T)$ was measured while cooling and warming in a field, respectively. Temperature and magnetic field dependence of thermopower and resistivity were measured simultaneously using a homemade setup.[11] Differential Thermal Analysis (DTA) during isothermal sweep was measured with a homemade set up which makes use of two Pt-1000 temperature sensors, with the sample on one of them while the second Pt resistor serves as a reference.[12] As the sample undergoes a phase transition in the presence of the applied magnetic field, its temperature changes and the temperature difference $\Delta T$ between the two sensors is measured.

The inset of Fig.1(a) shows $M(T)$ at $\mu_0 H = 0.1T$ in the ZFC, FCC and FCW modes. $M(T)$ varies very little with temperature until 40K in the ZFC mode, then decreases rapidly, goes through a dip around the Neel temperature $T_N = 120K$ and a peak at the charge-orbital ordering temperature $T_{CO} = 180K$. On the other hand, the FCC curve diverges from the ZFC curve below 40K and it continues to increase with lowering temperature. The rapid increase of FCC-$M(T)$ below 120K suggests that the samples undergoes a pseudo CE type antiferromagnetic state in which antiferromagnetically ordered Mn spins are canted. Main panel of Fig.1(a) shows $M(T)$ at $\mu_0 H = 1$ and 1.5T. The difference between ZFC and FCC magnetization is hardly recognizable at $\mu_0 H = 1T$ and the FCC and FCW curves exhibit a small hysteresis as like the data for 0.1T. In contrast, FCC and FCW curves show pronounced hysteresis in the temperature range $T = 40-120K$ under 1.5T and the ZFC curve diverges from the FCW-$M(T)$ below 120K. This difference suggests a dramatic change in the magnetic ground state when the field exceeds 1T. Fig. 1(b) shows $M(T)$ at $\mu_0 H = 2, 4, 5$ and 7T. While the ZFC curve shows a strong deviation from the FCW curves for $\mu_0 H = 2$ and 3T, the difference vanishes and hysteresis width decreases at 5 and 7T

Surprisingly, the temperature at which $M(T)$ rises rapidly during FCC shifts to higher temperature with increasing $H$ contrary to downward shift expected for an antiferromagnetic transition. It implies stabilization of ferromagnetic state at $\mu_0 H > 1$T. The ferromagnetic Curie temperature $T_C$, defined as the minimum in $dM/dT$ of the rapidly ascending part of FCC-$M(T)$ or descending part of FCW-$M(T)$ increases with increasing $H$. The field dependences of $T_C$ along with $T_{CO}$ are shown in the inset of fig. 1(b). The $T_{CO}$ decreases from 185K to 175K as $H$ changes from $\mu_0 H = 0.1$T to 5 T and $T_C$ increases from 49K for 170K for $\mu_0 H = 1.5$T to 7T. It suggests that the antiferromagnetic ground state transforms into a ferromagnetic state through a first-order phase transition above 1T and hence $T_C$ shifts up with increasing magnetic field.

Fig. 2(a) shows $M(H)$ isotherms at selected temperatures measured after cooling in zero field. During the first-field sweep ($\mu_0 H = 0 \rightarrow 7$T at 10K) marked as path 1, $M(H)$ initially increases linearly with $H$ as expected for an antiferromagnet and then shows an abrupt increase when the field exceeds a critical value $\mu_0 H_c = 4.6$T followed by a tendency to saturate at higher fields. We obtained saturation magnetization $M_s = 3.58$ $\mu_B$/f.u. at $\mu_o H = 7$T, which is closer to 3.5 $\mu_B$/f.u. expected for this composition and this small excess magnetic moment is possibly due to partial alignment of Nd-4f moment. Field cycling along 7T$\rightarrow$-7T (path 2) and then from -7T$\rightarrow$7T (path 3) produces a soft ferromagnetic hysteresis loop. Hence, $M$ in the initial field-sweep is irreversible. Similar irreversible behavior is seen in $M(H)$ up to 40K. Although $M(H)$ in initial field-sweep at $T = 60$ and 70K deviates from the subsequent field sweeps, $M(H)$ shows reversible metamagnetic transitions in path 2 and 3. The $M(H)$ loop in the first and subsequent field sweeps overlap on each other for $T \geq 60$K.

Fig. 2(b) shows results of DTA done at fixed base temperatures. During the first-field sweep ($0 \rightarrow +7$T) at 60K, the temperature difference $\Delta T$ between the two Pt sensors increases steadily and exhibits an exothermic peak ($\Delta T = +80$ mK) at a field corresponding to the metamagnetic transition ($\mu_0 H_c = 2.65$T) and then decreases. During field sweep H= +7T $\rightarrow$0T, $\Delta T$ exhibits an endothermic peak of smaller value ($\Delta T = -17$ mK) and another exothermic peak ($\Delta T = +24$ mK) when $H$ increases in negative direction. A mirror image is produced in the reverse sweep H= -7T →+7T. The important highlight is that the value of $\Delta T$ in the first cycle is larger than in subsequent field sweeps. This irreversibility in DTA clearly mimics the trend seen in $M(H)$. Similar behavior is seen at $T = 70$K but the value of $\Delta T$ peak is smaller than at $T = 60$K. At $T \geq 100$K, the irreversibility in $\Delta T$ vanishes and we again see clear anomalies in $\Delta T$ during the metamagnetic transitions. This $\Delta T$ indicates that heat is released from the sample to the surrounding during the field-induced transition from the antiferromagnetic to the ferromagnetic state and heat is absorbed during the reverse transition due to intrinsic magnetocaloric effect. The field-induced irreversibility in $M$ was reported in diverse materials such as phase separated manganites,[13] $CoFe_2$,[14] and $Mn_2PtGa$.[15] All these compounds undergo first-order magnetic phase transition in which the high temperature phase is supercooled to low temperature and coexists with the low temperature majority phase. We believe that the irreversibility found below 40K in $M(H)$ and DTA is caused by irreversible structural transition of the majority phase and metastable states formed with high energy barriers between them.

Fig.3(a) shows $\rho(T)$ under different magnetic fields. $\rho(T,H=0)$ increases with decreasing temperature and exceeds the instrument limit below 80K, whereas it exhibits insulator to metal transition at $T_{IM} = 96$K under 3T field. However, $\rho$ (10K, 3T) >$10^2$ $\Omega$ cm in the ZFC mode and it decreases with increasing temperature and merges with the FCW curve above 40K. Hysteresis between the FCC and FCW curves are notable. As H increases, $T_{IM}$ shifts up to 154 and 178K for $\mu_oH = 5$ and 7T, the irreversible nature decreases and the hysteresis narrows. Electrical conduction in paramagnetic state of manganites is generally dominated by adiabatic hopping of small polarons obeying the relation: $\rho(T) = \rho_0 T \exp(E_\rho / k_B T)$ where $E_\rho$ is the activation energy for the electrical transport and $k_B$ is the Boltzmann constant. $E_\rho = W_H + E_s$, where $E_s$ is the activation energy for thermopower and $W_H$ is the binding energy for small polaron. The inset of Fig. 3(a) shows the plots of $\ln(\rho/T)$ versus $1/T$ for $\mu_oH = 0$, 5 and 7T. All the curves show linear behavior at high temperature, which suggest that thermally activated polaronic conduction dominates the charge transport in the paramagnetic state. The high temperature data can be fitted with two straight lines with different slopes and the deviation occurs at $T_{CO}$. The slope of the straight line above $T_{CO}$ is less than the slope of second straight line below $T_{CO}$ and the estimated activation energy $E_\rho$ = 159.53 meV for $T > T_{CO}$ is less than $E_\rho$ = 182.14 meV for $T < T_{CO}$ in zero field. The activation energy above $T_{CO}$ decreases to 148.02 meV(124 meV) for $\mu_o H$ = 5T(7T).

Fig. 3(b) shows the temperature dependence of thermopower ($\alpha$) under the same magnetic fields used for $\rho(T)$. At $T$ = 350K, $\alpha$ = -11 µV/K and it gradually decreases in value with decreasing temperature, crosses over to positive values around 218K and reaches a maximum value of $\alpha$= +54 µV/K at $T$ = 82K below which it is not measurable due to very high resistance. It is not uncommon to find negative $\alpha$ in manganites for $T >> T_C$ and it arises due to contributions from spin entropy and the mixing entropy (Heikes term) in the expression for thermopower, discussed in earlier work. [16] When field-cooled under 3T, $\alpha(T)$ reaches a maximum value at $T_\alpha$ = 110K below which it decreases rapidly to zero and even changes sign (= -1µV/K) in the metallic state. It is to be noted that $\rho(T)$ shows the *I-M* transition in the same temperature range. The ZFC value of $\alpha$(10K, 3T) is higher than in the FCW mode but the irreversibility is less pronounced compared to $\rho(T)$. Further, FCC and FCW $\alpha(T)$ show hysteresis similar to $\rho(T)$. With increasing $H$, $T_\alpha$ shifts to higher temperature ($T_\alpha$ = 160 and 180K for 5T and 7T, respectively) and the hysteresis decreases. When $\mu_0 H$ = 7T, $\alpha(T)$ is positive only in narrow temperature region $T$ = 175-187K around the peak. A comparison with Fig. 3(b) suggest that the features observed in $\alpha(T)$ closely tracks $\rho(T)$ which itself tracks $M(T)$. When small polaron conduction dominates, $\alpha(T)$ is expected to follow the relation $\alpha = \frac{k_B}{e}\left(\frac{E_\alpha}{k_B T} + \alpha'\right)$ where $E_\alpha$ is the activation energy for thermopower and $\alpha'$ is related to the kinetic energy of polarons. If $\alpha' \ll 1$, means that the transport is due to small polaron hopping and $\alpha' > 2$ means large polaron hopping. [17] We show $\alpha(T)$ versus $1/T$ in the inset of Fig. 3(b) and $\alpha(T) \propto 1/T$ under all $H$ above $T_{CO}$ but a slope change at $T_{CO}$ is clearly observable in 0T. The calculated $E_\alpha$ are 11.93 (0T), 8.25 (5T) and 4.9 meV (7T) for $T > T_{CO}$ and 14.45 meV (0T) for $T < T_{CO}$.

The calculated is $\alpha' = 0.463$ in zero field and $E_\alpha \ll E_\rho$ means hopping of small polarons dominates the electrical conduction.

Fig. 4(a) and (b) shows $\rho(H)$ and $\alpha(H)$ at selected temperatures. When $T = 10K$, $\rho(H)$ during the initial field-sweep is not measurable for $\mu_0 H < 4.2T$, but $\rho(H)$ shows a rapid decrease between 4 and 5T followed by a slow variation at still higher fields. As the field is cycled (+7T→-7T→+7T), resistivity remains at lower values than during the first-sweep. A similar trend persists up to $T = 40K$. When $T = 60K$, $\rho(H)$ drops precipitously from ~$10^5$ Ω cm to a few mΩ cm in the initial field-sweep around 2.65T. Subsequently, as the field is cycled, $\rho(H)$ shows a hysteresis loop that is reversible in both the directions of $H$. At $T = 100$ and 125K, $\rho(H)$ initially decreases gradually with increasing field and shows a rapid decrease during the field-induced metamagnetic transition and shows hysteresis when the field is reduced. The field dependence of $\alpha$, shown in Fig. 4(b) closely resembles $\rho(H)$. At $T = 10$ and 40K, $\alpha(H)$ is irreversible between the first forward field sweep (0 to +7T) and subsequent field sweeps. While $\alpha(H)$ shows reversible hysteresis during the field-induced metamagnetic transition at $T = 100$, 125 and 150K but it is absent for $T \geq 175K$.

Fig. 5(a) shows $\alpha(H)$ and $\rho(H)$ at $T = 70K$, on the left and right scales, respectively. As the field is increased, $\alpha(H)$ decreases abruptly around $H_c = 3T$ and that is accompanied by an abrupt decrease in $\rho(H)$. Both $\alpha(H)$ and $\rho(H)$ trace a nice hysteresis loop but of smaller magnitudes in subsequent field cycle. Both $\alpha(H)$ and $\rho(H)$ at 125 K exhibits reversible hysteresis loop and show similar field dependences as shown in Fig.5(b). Fig. 5(c) shows the magnetoresistance MR= $[\rho(H)-\rho(0)]/\rho(0)$ at $T = 100, 125, 150, 175$, and 200K. The MR at 100 K is negligible at lower fields when the sample is in the AFM state. As $H$ increases, the AFM phase partially collapses and ferromagnetic phase evolves during which MR increases in magnitude abruptly by 100%. At higher fields, MR varies gradually when the sample is fully converted into FM metallic phase. The critical field for the metamagnetic transition shifts to higher field with increasing temperature and it is not observed for $T>175K$. The magnetothermopower,

MTEP = $\Delta\alpha/\alpha$ = [$\alpha(H)-\alpha(0)$]/$\alpha(0)$] reaches -100% at 4.4(3.3)T and $T$ = 125 (100)K. In fact, MTEP slightly exceeds 100% because $\alpha$ becomes negative. The MTEP hardly varies with $H$ after the metamagnetic transition is completed. The rapid increase in the magnitude of MTEP shifts to higher field as $T$ increases. However, while MR is only -51% and -35% at $T$ = 175 and 200K, MTEP reaches -100% at these temperatures. We are not aware of any theoretical model which predicts $\alpha(H)$. We find $\alpha(H)$ is linearly proportional to the $M^2$ at 200K and 225K as shown in the inset when the sample is in the paramagnetic state but deviates from the linearity at 175K and below(not shown here).

The data presented above suggests that a change in the magnetic state of the sample has a dramatic effect on $\alpha$(measured in an open circuit with no net current flow) and it resembles the effect of magnetic field on the resistivity measured in presence of a forced current. In our sample, $\alpha(H=0T)$ is positive and large at 80K when the neighboring Mn spins are antiparallel and turns to small and negative value when the neighboring spins are forced to become parallel at a high magnetic field which implies spin-dependent thermal transport. The observed MTEP is much higher than $\Delta\alpha/\alpha$ ($H$=0) = 65.3% for $\Delta H$ = 8T found in $La_{0.7}Ca_{0.3}MnO_3$ thin film.[5,6] The charge carriers (holes) in the title composition are localized and ordered in the AFM state. As the AFM state collapses and transforms into ferromagnetic with increasing $H$ at a constant temperature, charge ordering is destroyed and holes become delocalized leading to metallic conduction. Under the imposed temperature gradient, delocalized holes diffuse from the hot to the cold end which results in the observed MTEP. However, the observed MTEP is much larger than MR at higher temperatures, e.g., at $T$ = 175 and 225K when the sample is in paramagnetic state. A discrepancy between the magnitudes of MR and MTEP could arise due to the fact that while the electrical conductivity is proportional to the density of states at the Fermi level, thermopower probes the asymmetry in the density of states at the Fermi level.[7,18] The larger the asymmetry in the density of states for spin and spin down holes in the zero-field, larger could be MTEP. As the field increases, Zeeman splitting between the spin-up and spin down-density of states will also increase. The change of sign from positive to negative thermopower at high fields suggests that electron filling at the Fermi level of one of the spin split bands increases with increasing magnetic field.

In summary, we found that field-induced collapse of antiferromagnetic state in $Nd_{0.75}Na_{0.25}MnO_3$ at a fixed temperature or as a function of temperature is accompanied by giant negative magnetoresistance and magnetothermopower as high as 100% below 200K. Magnetothermopower shows an irreversibility between the first and subsequent field sweeps for T< 40K which corroborates with magnetization. In addition, irreversible heat evolution at low temperature was also demonstrated using the differential thermal analysis. Our results suggest a close interplay between magnetization, electrical resistance and heat transport in the studied sample. The exact origins of sign change in thermopower and MTEP larger than MR have to be understood.

**Figure captions:**

**Fig.1:** Magnetization(*M*) as a function of temperature for $\mu_0H$ = 1, 1.5 T(a) and 2, 3, 5 and 7T(b). Inset 1(a) shows *M(T)* under $\mu_0H$ = 0.1T and inset 1(b) shows field dependence of the charge ordering temperature $T_{CO}$ and the ferromagnetic Curie temperature $T_C$ while cooling (open symbol) and warming (closed symbol). See the text for explanations for other notations.

**Fig.2:** Isothermal field dependence of (a) magnetization and (b) temperature change($\Delta T$) of the sample measured using the DTA technique.

**Fig.3:** Temperature dependence of (a) resistivity ($\rho$) and (b) thermopower ($\alpha$) for different *H*. The inset of fig. 3(a) and 3(b) shows polaronic fit for $\ln(\rho/T)$ vs $1/T$ and $\alpha$ vs $1/T$, respectively.

**Fig.4:** Field dependence of (a) dc resistivity($\rho$) and (b) thermopower($\alpha$) at selected temperatures.

**Fig.5:** $\rho(H)$ and $\alpha(H)$ at (a)*T* = 70K and (b)125K, (c) magnetoresistance and (d) magnetothermopower at selected temperatures. Inset in fig.(d) shows dependence of thermopower on square of the magnetization in the paramagnetic state.

**References**

[1] J. Shi, K. Petit, E. Kita, S.S.P. Parkin, R. Natakani and M. Salamon, Phys. Rev.B **54**, 15273 (1996).

[2] L. Gravier, A. Fabian, A. Rudolf, A. Cachin, J.-E. Wegrowe, and J.-P. Ansermet, J. Magn. Magn. Mater. **271**, 153 (2004).

[3] M. Walter, J. Walowski, V. Zbarsky, M. Munzenberg *et al.*, Nature Mater. DOI:10.1038/NMAT3076 (2011).

[4] W. Li, M. Hehn, L. Chaput, B. Negulescu, S. Andrieu, F. Montaigne and S. Mangin, Nature Commun. **3**.744 doi:10.1038/ncomms1748 (2012).

[5] M. Jaime, M. B. Salamon, K. Petit, M. Rubinstein, R. E. Treece, J. S. Horwitz, and D. B. Chrisey, Appl. Phys. Lett. **68**, 11 (1996).

[6] B. Chen, C. Uher, D. T. Morelli , J. V. Mantese, A. M. Mance, and A. L. Micheli, Phys. Rev. B **53**, 5094 (1996).

[7] A. Asamistu, Y. Moritomo, and Y. Tokura, Phys. Rev B **53,** 2952 (1996).

[8] V.H. Crespi, Li Lu, T. X. Xia, K. Khazani, A. Zettel, and Marvin Cohen, Phys. Rev. **53**, 14303 (1996).

[9] S. Pal, A. Banerjee, E. Rozenberg and B.K. Chaudhuri, J. Appl. Phys. **89,** 4955 (2001).

[10] B. Samantaray, S. Ravi, A. Das and S.K. Srivastava, J. Appl. Phys. **110**, 093906 (2011).

[11] D. V. Maheswar Repaka, T. S. Tripathi, M. Aparnadevi, and R. Mahendiran, J. Appl. Phys. **112,** 123915 (2012).

[12] M. Quintero, J. Sacanell, L. Ghivelder, A.M. Gomes, A.G. Leyva, F. Parisi, Appl. Phys. Lett. **97**, 121916 (2010); M. Aparnadevi, S. K. Barik and R. Mahendiran, J. Magn. Magn. Mater. **324**, 3351 (2012).

[13] R. Mahendiran, A. Maignan, S. Hébert, C. Martin, M. Hervieu, B. Raveau, J. F Mitchell, and P. Schiffer, Phys. Rev. Letts, **89** 286602 (2002) .

[14] V. K. Sharma, M. K. Chattopadhyay, and S. B. Roy, Phys. Rev. B **76**, 140401(R) (2007).

[15] A. K. Nayak, M. Nicklas, S. Chadov, C. Shekar, Y. Skourski, J. Winterlik and C. Felser, Phys. Rev. Lett. **110**, 127204 (2013).

[16] M.F. Hundley and J.J. Neumeier, Phys. Rev. B **55**, 11511 (1997).

[17] K. Sega, Y. Kuroda and H. Sakata, J. Mater. Sci. **33**, 1303 (1998).

[18] M. Czerner, M. Bachmann and C. Heiliger, Phys. Rev. B **83**, 132405(2011).

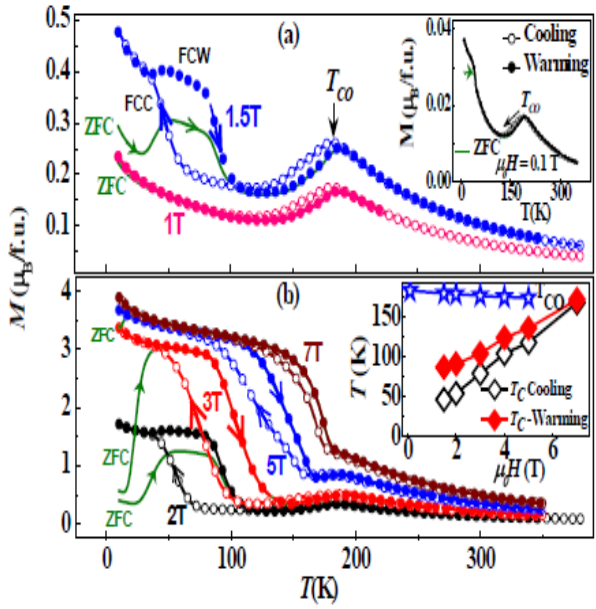

Fig. 1. D. V. Maheswar Repaka *et al*.

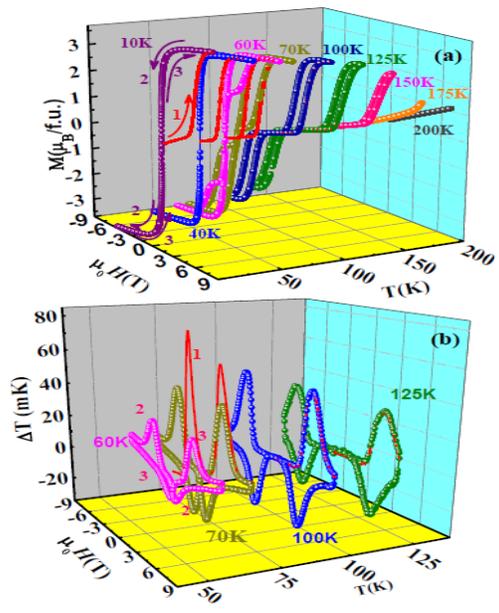

Fig.2 D. V. Maheswar Repaka *et al*.

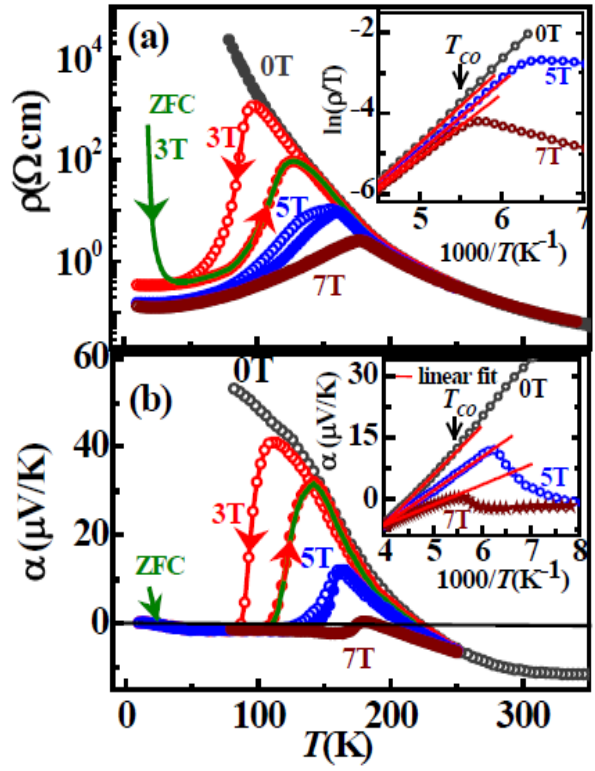

Fig. 3. D. V. Maheswar Repaka *et al.*

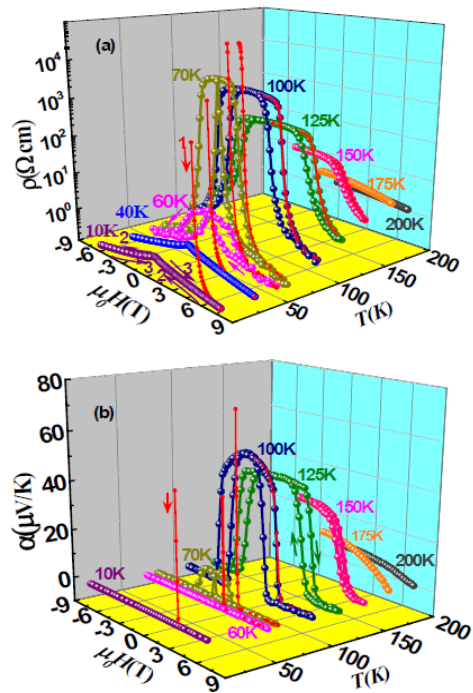

Fig. 4  D. V. Maheswar Repaka *et al*.

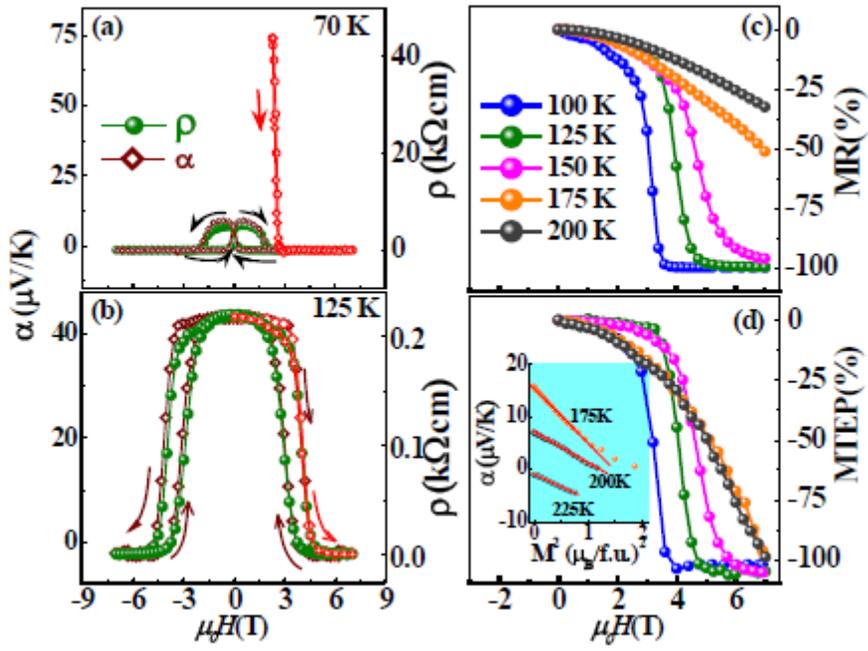

Fig. 5.   D. V. Mahehswar Repaka *et al*.